# Cold + Hot Dark Matter and
# the Cosmic Microwave Background

Scott Dodelson

NASA/Fermilab Astrophysics Center

Fermi National Accelerator Laboratory, Batavia, IL  60510-0500

Evalyn Gates

Department of Astronomy & Astrophysics

Enrico Fermi Institute, The University of Chicago, Chicago, IL  60637-1433
and
NASA/Fermilab Astrophysics Center

Fermi National Accelerator Laboratory, Batavia, IL  60510-0500

Albert Stebbins

NASA/Fermilab Astrophysics Center

Fermi National Accelerator Laboratory, Batavia, IL  60510-0500

## ABSTRACT

We examine the cosmic microwave background power spectrum for adiabatic models with a massive neutrino component. We present the results of a detailed numerical evolution of cold + hot dark matter (CHDM) models and compare with the standard cold dark matter (CDM) spectrum. The difference is of order $5 - 10\%$ for $400 < l < 1000$ for currently popular CHDM models. With semi-analytic approximations, we also discuss the relevant physics involved. Finally we remark on the ability of future experiments to differentiate between these models. An all-sky experiment with a beam size smaller than 30 arcminutes can distinguish between CHDM and CDM if other cosmological parameters are known. Even allowing other parameters to vary, it may be possible to distinguish CDM from CHDM.

## 1. Introduction

Since the discovery of the anisotropies in the cosmic microwave background radiation (CMB) by the COBE satellite (Smoot et al. 1992), experiments designed to probe the spectrum of these anisotropies at increasingly smaller angular scales continue to grow in number. Ultimately, these experiments will be able to distinguish amongst cosmological models (Hinshaw, Bennett, & Kogut 1995; Knox 1995; Jungman et al. 1995). Of particular interest here is the class of models (Shafi & Stecker 1984) with both a cold dark matter component and a neutrino with a mass of order $1 - 10$ eV. These cold plus hot dark matter (CHDM) models have been shown (van Dalen & Schaefer 1992; Davis, Summers, & Schlegel 1992; Klypin et al. 1993) to reproduce the observed large scale structure more successfully than cold dark matter (CDM), so such models have been extensively studied recently. The question naturally arises then as to whether the upcoming generation of CMB experiments will be able to distinguish CDM from CHDM. This is a fairly complicated question, depending in large part on the nature of the experiments. Clearly, though, the first step is to calculate the spectrum of anisotropies in CHDM and compare it to the CDM spectrum. This we do here (see also Holtzman 1989; Ma & Bertschinger 1995; Holtzman, Klypin & Primack 1995). In § 2 we present our results along with a description of the code which produced them.

The difference between CDM and CHDM is fairly small, so one might think it would be difficult to try to understand the physical reasons for these small differences. Within the last year, however, there has been a great advance in our ability to explain features in CMB spectra. We refer to the work of Hu & Sugiyama (1995) (hereafter referred to as HS), who have greatly improved upon the accuracy of semi-analytic methods for calculating the CMB anisotropy spectrum, and thereby elucidated many of the subtleties which lead to the predicted spectrum. Using these methods we will try to isolate the main causes responsible for the difference between these two models.

Given these differences, there exists the tantalizing possibility of using CMB measurements to distinguish between the two models. In § 4, we address this possibility and argue that if all other cosmological parameters are known, *CMB measurements can distinguish between CDM and CHDM.* Even if some of the other cosmological parameters are allowed to vary, an experiment with a small beam ($\lesssim 20$ arcminutes) will be able to distinguish CDM from CHDM.

Throughout this paper we will present our results in the language of $C_l$'s. If the temperature on the sky is expanded as $T(\theta, \phi) = \sum_{lm} a_{lm} Y_{lm}(\theta, \phi)$, then the expected value of the square of the coefficients is defined as

$$C_l \equiv \langle |a_{lm}|^2 \rangle. \qquad (1)$$

Large $l$ corresponds to small angles. For reference, the COBE experiment was sensitive to multipoles $l \lesssim 30$.

## 2. Numerical Results

To get the $C_l$'s in a given model one must solve the coupled Einstein-Boltzmann equations for the perturbations to the smooth homogeneous background. Since the equations are coupled, even if we are interested only in the CMB anisotropies today, we still must follow the perturbations to all other species: the CDM component; the massless neutrinos; the metric; the baryons; and finally the massive neutrino. These equations and the numerical techniques used to solve them have been presented many times over the last twenty five years (e.g. Peebles & Yu 1970; Wilson & Silk 1980; Bond & Efstathiou 1984; Vittorio & Silk 1984; Gouda & Sugiyama 1992; Dodelson & Jubas 1993; Hu, Scott, Sugiyama & White 1995). This work has been done primarily assuming neutrinos are massless. Here we focus on the changes needed to treat massive neutrinos (see also Ma & Bertschinger 1995).

To describe the differences between CDM and CHDM, let us write down the equation governing the perturbation to a collisionless massive species:

$$\frac{\partial \Theta}{\partial \eta} + \frac{\vec{q}}{E} \cdot \vec{\nabla} \Theta = S(\vec{x}, t) \qquad (2)$$

where $\eta$ is the conformal time defined in terms of the scale factor $a$, $\eta \equiv \int_0^t dt'/a(t')$; $\Theta$ is essentially the deviation of the distribution function from its zero order value; the comoving momentum is $\vec{q}$; $E = \sqrt{q^2 + (ma)^2}$; and $S$ is some source term proportional to the metric perturbations, i.e. the gravitational fields. In principle $\Theta$ is a function of seven variables: three spatial, three momentum, and one time. For almost all models though, there are symmetries which drastically reduce the number of dependent variables. In models of scalar (density) perturbations the Fourier transform of $\Theta(\vec{x}, \vec{q}, t)$, call it $\tilde{\Theta}(\vec{k}, \vec{q}, t)$, depends only on $(k, q, \vec{k} \cdot \vec{q}, t)$. Since the equations are linear, each $k-$ mode evolves independently; to deal with the $\vec{k} \cdot \vec{q}$ dependence, one usually expands $\tilde{\Theta}$ in a series of Legendre polynomials: $\tilde{\Theta} = \sum_l (2l+1)(-i)^l P_l(\vec{k} \cdot \vec{q}) \tilde{\Theta}_l$. Then equation 2 becomes

$$\frac{\partial \tilde{\Theta}_l}{\partial \eta} + \frac{q}{E} k \left[ \frac{l\tilde{\Theta}_{l-1} + (l+1)\tilde{\Theta}_{l+1}}{2l+1} \right] = \tilde{S}(k, t) \qquad (3)$$

This equation is very useful; although it has the disadvantage of being in Fourier space which is harder to visualize, understanding the difference between cold and hot particles or between massive and massless particles is particularly easy starting from equation 3.

For a cold component, i.e. a component whose velocity dispersion is negligible, the distribution function can be described solely in terms of a density and a velocity, given by $\Theta_0$ and $\Theta_1$ respectively. The distribution function is thus given by 2 components, whose evolution is described by two 1st order equations which can be derived from equation 3. One may even choose the coordinate frame such that the velocity is always zero, leaving only the density and effectively dropping all modes with $l > 0$. This is a great simplification. On the other hand, for a massless component [such as massless neutrinos], the different $l$−modes *do* talk to one other. If an initial perturbation is set up in the $l = 0$ mode, as time evolves the higher $l$−modes will also become non-zero. At very early times, $k\eta \ll 1$ so the second term [$\sim k\Theta$] in equation 3 is negligible compared to the first [$\sim \Theta/\eta$]. Only when $k\eta$ becomes of order unity will a perturbation in the $l = 0$ mode *freestream* into the $l = 1$ mode. Thereafter, each subsequent $l$−mode gets populated at $\eta \sim l/k$. Numerically, we account for freestreaming by gradually adding more and more $l$−modes to our hierarchy of equations as time evolves. Keeping track of all the relevant $l$−modes is the complicated part of solving equation 3 for massless particles; the simple part is that the equation is momentum independent. This follows since in the massless case, $q/E = 1$. Thus we need solve equation 3 only once; this gives $\tilde{\Theta}(q)$ for all $q$. This simple feature is lost when we pass to the massive neutrino case.

For massive neutrinos, $q/E \neq 1$ and we must solve equation 3 for many values of $q$. We use a grid with 32 values of q, spaced properly for Gauss-Legendre integration, as proposed by Bond & Szalay (1983). If massive neutrinos freestreamed in the same manner as massless ones, then we would be in trouble. Keeping track of all the different massive neutrino momentum modes would slow down the program by at least a factor of 30. However, physics helps out: massive neutrinos stop freestreaming because a given $q$− mode gets more and more non-relativistic as time evolves. It is easy to see this from equation 3. The first term is of order $\Theta/\eta$; the second is of order $k\Theta q/E = k\Theta/\sqrt{1 + (a/a_{NR})^2}$, where the scale factor at which a given $q$− mode becomes non-relativistic is

$$a_{NR} \equiv q/m. \tag{4}$$

We have seen that freestreaming occurs when the second term is of order the first; that is, when the scale of the perturbation $\sim 1/k$ is of order the freestreaming scale:

$$\frac{1}{k} \sim \frac{1}{k_{\text{freestream}}} \equiv \frac{\eta}{\sqrt{1 + (a/a_{NR})^2}}. \tag{5}$$

At early times, the square root here is unity, so we return to the criterion of the massless case. Physically this is reasonable: at early times, even massive neutrinos are very relativistic. At late times though the right hand side goes as $\eta/a \sim 1/\eta$, so it will eventually become smaller than the left side. This reflects the fact that as time evolves the massive neutrinos become increasingly non-relativistic and are no longer able to freestream. For any given momentum mode, then, there are three separate regimes: (i) perturbation outside horizon ($k\eta \lesssim 1$) – no freestreaming (ii) perturbation enters horizon ($k\eta \gtrsim 1$) – freestreaming and (iii) perturbation size becomes larger than freestreaming scale ($k\eta < a/a_{NR}$) – no freestreaming. These three regimes are illustrated in Figure 1 for several different momentum modes. Numerically, we can stop adding new $l-$modes once we pass into region (iii). Even better, for the purposes of CMB anisotropies, only the $l = 0, 1$ neutrino modes couple [indirectly] to the photons. Once we pass into region (iii), these get less and less contaminated from higher $l$−modes. So we can start dropping these higher modes, thereby reducing the number of equations that need to be solved.

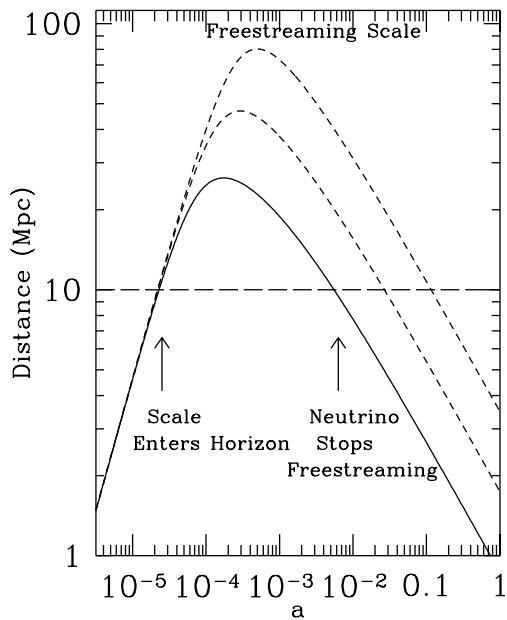

Fig. 1.— Massive neutrino freestreaming. The freestreaming scale is shown here for the average momentum mode ($q = 3T$; solid line) and some higher momentum modes ($q = 6, 12T$; dashed lines). At early times there is no freestreaming since the scale of the perturbation is larger than the horizon. At intermediate times, the perturbation enters the horizon and freestreams. At late times, the neutrino becomes non-relativistic and stops freestreaming. The scale of the perturbation is constant in comoving coordinates, $2\pi/k$. Here we show one specific value of $k$ (long dashed horizontal line); this scale is so small that it contributes very little to anisotropies for $l \lesssim 1000$. The more relevant scales will experience even less freestreaming.

The results of our numerical integration are shown in figure 2. For comparison we present also a *standard* cold dark matter spectrum where the neutrinos are assumed massless. For ease of comparison, except where otherwise stated, we will only consider $\Omega_0 = 1$, $\Omega_b = 0.05$, $H_0 = 50 \mathrm{km/sec/Mpc}$ with a primordial Harrison-Zel'dovich spectrum for all models. Note that here and throughout, the theories have been normalized to be equivalent at low $l$. In a CHDM model with $\Omega_\nu = 0.2$ and one massive neutrino species, the spectrum deviates (at about the $5 - 10\%$ level) from the standard CDM spectrum for $l \gtrsim 400$, as shown in figure 2. Increasing $\Omega_\nu$ to 0.3 increases the departure from the standard CDM spectrum, although only slightly. Increasing the number of massive neutrino species to two (for simplicity we assume degenerate neutrino masses) in a CHDM model with $\Omega_\nu = 0.2$ further shifts the position of the peaks relative to the standard CDM spectrum.

## 3. Discussion and semi-analytic approximation

A careful inspection of figure 2 reveals two effects. First the amplitude of the CHDM perturbations are larger than those of CDM spectra. Second the oscillations in the amplitude are shifted systematically to the left, i.e. towards smaller $l$. We will try to explain the reason for these small differences. As we shall see the "horizontal shift" is explained by one effect while the "vertical shift" is the result of several different effects. Below, in §3.2, we will discuss the horizontal shift first as we will find it easier to discuss the vertical shift only after "correcting" for the horizontal shift. In § 3.3 we discuss the vertical shift.

### 3.1. Expansion Law and Horizons

During the epochs where the anisotropies are generated the neutrinos only have significant interaction with the photons via gravity. Thus differences between CHDM and CDM must be due to the differences in the gravitational field of the neutrinos in the two models. Conceptually we can divide this gravitational field difference into two parts. The homogeneous component of the gravitational field determines how the difference in the equation of state for a massive and massless neutrinos affects the expansion law of the universe. The inhomogeneous component of the gravitational field, determined by the difference in how massive and massless neutrinos cluster when gravitational inhomogeneities are present, affects the growth of fluctuations in other components. We shall see that there are a variety of ways in which varying the homogeneous and inhomogeneous components of the gravitational field of the neutrinos can effect the CMB anisotropies. Here we compute the difference in the expansion law and later discuss how this effects anisotropies.

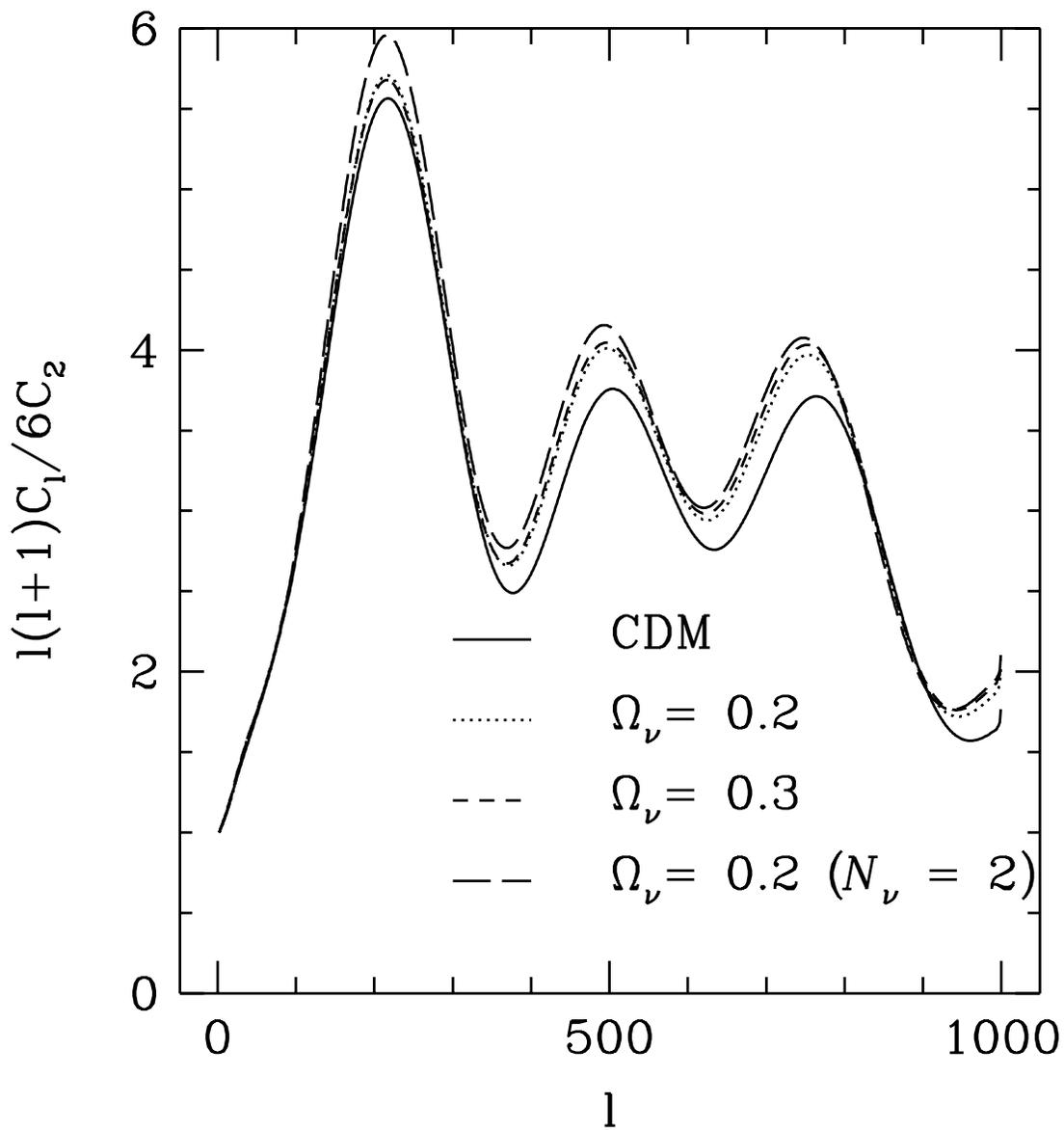

Fig. 2.— CMB power spectrum in CDM and three variants of CHDM.

The Friedmann equation tells us that the relation between the conformal time ($\equiv$ comoving causal horizon) and the scale factor is fully determined by the run of the average cosmological density with redshift. In particular, assuming $\Omega_0 = 1$, and using units such that $a_0 = 1$, we find

$$\eta(a) = \int_0^a \frac{da'}{\dot{a}'} = \eta_0 \frac{\int_0^a \frac{da'}{a'^2} \sqrt{\frac{\rho_0}{\rho(a')}}}{\int_0^1 \frac{da'}{a'^2} \sqrt{\frac{\rho_0}{\rho(a')}}} \tag{6}$$

where $\rho_0$ is the mean cosmological density today. This gives the causal horizon as a function of the scale factor. The expansion law, $a(\eta)$, is just the inverse of this function. For $a < 1$, the denominator is less sensitive to changes in cosmological parameters than is the numerator.

For CMB anisotropies, another important horizon is the comoving distance sound in the photon-baryon fluid would have traveled, which is given by

$$r_{\rm s}(a) = \int_0^{\eta(a)} d\eta' \, c_{\rm s}(a') = \eta_0 \frac{\int_0^a \frac{da'}{a'^2} c_{\rm s}(a') \sqrt{\frac{\rho_0}{\rho(a')}}}{\int_0^1 \frac{da'}{a'^2} \sqrt{\frac{\rho_0}{\rho(a')}}} \tag{7}$$

The sound speed can be expressed in terms of the pressures and densities of the baryons and photons:

$$c_{\rm s}^2(a) = \frac{\partial p_{\gamma \rm b}}{\partial \rho_{\gamma \rm b}} = \frac{4p_\gamma + 4p_{\rm b}}{4\rho_\gamma + 3\rho_{\rm b}} \approx \frac{4p_\gamma}{4\rho_\gamma + 3\rho_{\rm b}} = \frac{4p_{\gamma 0}}{4\rho_{\gamma 0} + 3a\rho_{\rm b0}} = \frac{1}{3} \frac{1}{1 + \frac{3}{4}a\frac{\Omega_{\rm b0}}{\Omega_{\gamma 0}}} \tag{8}$$

where $\Omega_{\rm b0}$ and $\Omega_{\gamma 0}$ give the fraction of the critical densities in baryons and photons today. These two quantities are the same in the CHDM and CDM models we are comparing and hence the run of sound speed with $a$ is the same in the two models. The functions $r_{\rm s}(a)$ will be different between the two models since $\eta(a)$ is different and in particular because $\rho(a)$ will be different. Note that for many purposes it is easier to compare the CHDM and CDM models at the same $a$ when many quantities of interest, such as the free electron density, $\rho_\gamma$, and $\rho_b$, have the same value in both models, rather than at the same $\eta$, when almost all quantities are different. Note also that equations 7-8 become meaningless after the photons and baryons decouple at recombination.

In the CDM model we are considering the matter is always divided between non-relativistic and ultra-relativistic (effectively massless) species, i.e.

$$\frac{\rho}{\rho_0} = \frac{\Omega_{n0}}{a^3} + \frac{\Omega_{u0}}{a^4} \qquad \Omega_0 = \Omega_{n0} + \Omega_{u0} = 1 \tag{9}$$

where $\Omega_{n0}$ and $\Omega_{u0}$ give the fractional density of non-relativistic and ultra-relativistic species, respectively. In the CDM model, baryons and CDM particles constitute the non-relativistic species today, while the ultra-relativistic density is made up of the (massless) neutrinos and photons. In this model one can solve for the causal horizon and the sound horizon, $r_s$, analytically, giving the dependence of $r_s$ on the cosmological parameters $H_0$ and $\Omega_b$. Unfortunately no such analytical formulae giving the dependence on neutrino mass can be found in the CHDM model .

In the CHDM model, the energy density cannot be so neatly divided into non-relativistic and ultrarelativistic species. During the epochs of interest here, one or more of the neutrinos is undergoing the transition from ultra-relativistic to non-relativistic. We can express this in an equation similar to equation 9

$$\frac{\rho}{\rho_0} = \frac{\Omega_{b0} + \Omega_{c0}}{a^3} + \frac{\Omega_{\gamma0}}{a^4} \left( 1 + \frac{7}{8} \left( \frac{4}{11} \right)^{\frac{4}{3}} \left( (3 - \mathcal{N}_\nu) + \mathcal{N}_\nu\, F \left( a \left( \frac{11}{4} \right)^{\frac{1}{3}} \frac{m_\nu c^2}{k_B T_{\gamma0}} \right) \right) \right) \tag{10}$$

where we have assumed that $\mathcal{N}_\nu$ of the neutrino species have a degenerate mass, $m_\nu$, $\Omega_{c0}$ is the fractional density in CDM, and $F$ is an integral over the Fermi-Dirac distribution function:

$$F(y) = \frac{\int_0^\infty dx\, x^2 \sqrt{x^2 + y^2}\, (e^x + 1)^{-1}}{\int_0^\infty dx\, x^3\, (e^x + 1)^{-1}}. \tag{11}$$

As mentioned above, $\Omega_{\gamma0}$ and $\Omega_{b0}$ are the same in both models, while in the CHDM models the massive neutrinos contribute to $\Omega_{n0}$. Thus the density in CDM particles, $\Omega_{c0}$, in CHDM models must be somewhat smaller in order to maintain $\Omega_0 = 1$. In figure 3 we plot the ratio of the total density from equation 10 to that for the CDM model from equation 9 and show the dependence of this ratio on various cosmological parameters. In both the CHDM and CDM models the present density is fixed to the same value by the Hubble constant. Similarly at very large redshifts, when the particle masses contribute negligibly to the density, the density in both models is fixed by the measured temperature of the CMB, $T_{\gamma0}$. However, there is an intermediate epoch where the massive neutrinos are relativistic and no longer contribute to the density in non-relativistic particles $\Omega_{NR}$, while $\Omega_{NR}$ itself is still a significant fraction of the total density. Thus there is a narrow range of redshifts during which the $\rho(a)$ in the two models will disagree at the few percent level. The fact that the density is lower in the CHDM model is partly a result of the lower value $\Omega_{c0}$.

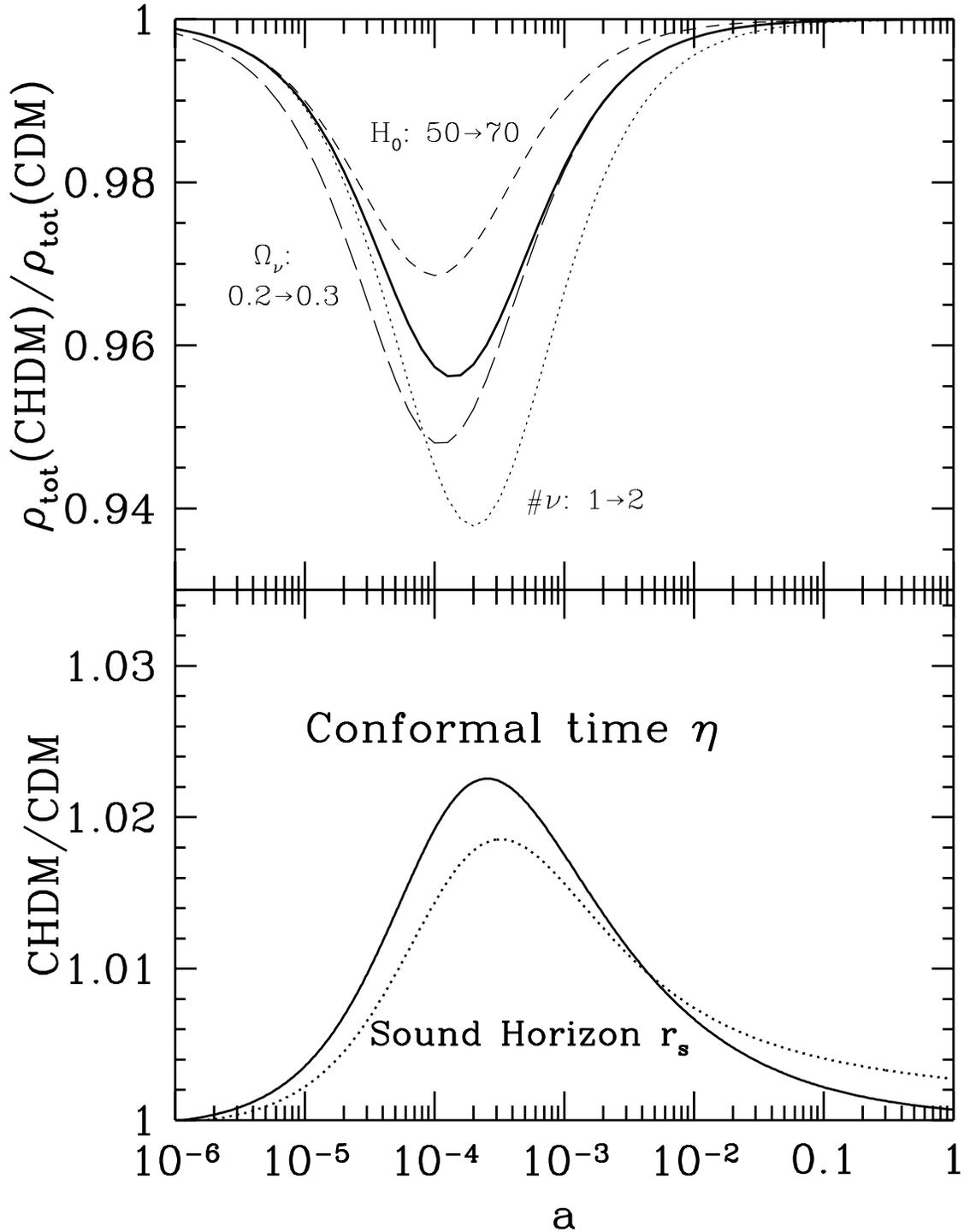

Fig. 3.— Plotted as a function of scale factor is the ratio of the total density in the CHDM model at a given scale factor to the total density in the CDM model at the same scale factor. The thick curve is for our canonical CHDM model, with $H_0 = 50$ km/sec/Mpc, one massive neutrino species, and $\Omega_\nu = 0.2$, while in each of the other curves one of these parameters has been changed as indicated. The bottom panel shows the ratio of the comoving horizons (both causal and sound) in the canonical CHDM model to those in the CDM model . The slight increase in the sound horizon at recombination ($a \sim 10^{-3}$) shifts the Doppler peaks in the CHDM model to lower $l$ than in the CDM model.

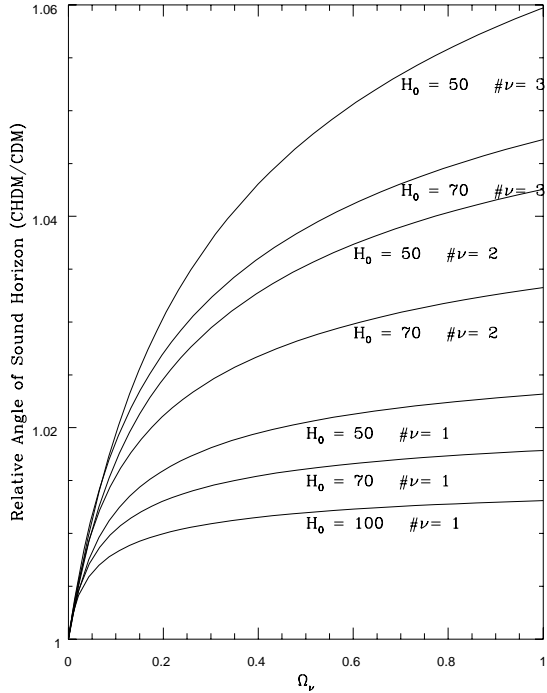

Fig. 4.— The shift in the sound horizon distance for various CHDM models as a function of the fraction of the critical density taken up by the neutrinos, $\Omega_\nu$, which also determine the neutrino mass. The models assume $\#_\nu$ of the neutrinos have masses assumed degenerate. Note that models with more than one massive neutrino have particularly large shifts. A larger Hubble constant leads to a somewhat smaller shift. The angular wavenumber of the 2nd and 3rd Doppler peaks are inversely proportional to the sound horizon.

From figure 3 we see that the densities are less in the CHDM model at the same redshift. This slows the expansion rate and allows more time for both light and sound waves to propagate further. This is evident from equations 6-7 where a smaller $\rho$ leads to an increase in $\eta$ and $r_s$. Note that the $\rho$-dependence in the denominators of these expressions is weaker than that for the numerator as long as $a < 1$. In figure 3 we have also plotted the fractional increase in the causal and sound horizon in our standard CHDM model over our standard CDM model. Changes in cosmological parameters will increase or decrease this difference as indicated by figure 4.

## 3.2. Horizontal Shift

The importance of the sound horizon derives from the fact that the oscillations in the $C_l$'s reflect the varying temporal phases of acoustic oscillations as a function of wavenumber. Before recombination, gravitational infall into potential wells is opposed by photon pressure, setting up acoustic oscillations. The density and velocity of the acoustic waves vary as

$$\delta(\eta) \propto \Theta_0 \approx \cos kr_s(\eta) \qquad v(\eta) \propto \Theta_1 \approx \sin kr_s(\eta) \tag{12}$$

in Newtonian gauge (see e.g. HS). Since the adiabatic (density) anisotropies are larger than the Doppler (velocity) anisotropies, and since there is a rough correspondence between spatial frequency, $k$, and angular frequency, $l$, one finds the oscillatory behavior for the $C_l$'s illustrated in figure 2. The peaks are located at $k \simeq n\pi/r_s$, for integer $n$. By increasing $r_s$ one therefore decreases the $k$ of a given temporal phase and hence shifts the peaks and the troughs to lower $l$. The angle subtended by the peaks is proportional to $r_s$ and in figure 4 we show how much the fractional increase in $r_s$ depends on various neutrino and cosmological parameters.

Now that we understand that the features in $C_l$ spectrum will be shifted in $l$ between CDM and CHDM, we can isolate the differences not explained by the shift by first shifting the CHDM spectrum in $l$−space and then noting what differences remain with the CDM spectrum. Figure 5 shows both the unshifted and the shifted comparison. From figure 5 we see that once the shift is removed, the fundamental difference between CDM and CHDM is a $\sim 5 - 10\%$ higher amplitude in CHDM. We now turn to this vertical shift.

## 3.3. Vertical Shift

In order to understand why the amplitude of the CHDM spectrum is 5 - 10% higher than the CDM spectrum (after shifting in $l$-space as described in the preceding section) we first consider the tight-coupling regime. As described in HS the equations detailing the evolution of the perturbations can be well approximated by an equation for the evolution of a tightly coupled photon-baryon fluid in a gravitational field. Following their prescription we have:

$$\ddot{\Theta}_0 + \frac{\dot{a}}{a}\frac{R}{1+R}\dot{\Theta}_0 + k^2 c_s^2 \Theta_0 = F(\eta) \tag{13}$$

where the forcing function $F(\eta) = -\ddot{\Phi} - \frac{\dot{a}}{a}\frac{R}{1+R}\dot{\Phi} - \frac{k^2}{3}\Psi$ is a function of the gravitational potentials $\Phi$ and $\Psi$. In the limit that the universe is matter-dominated $\Phi = -\Psi$ is the usual

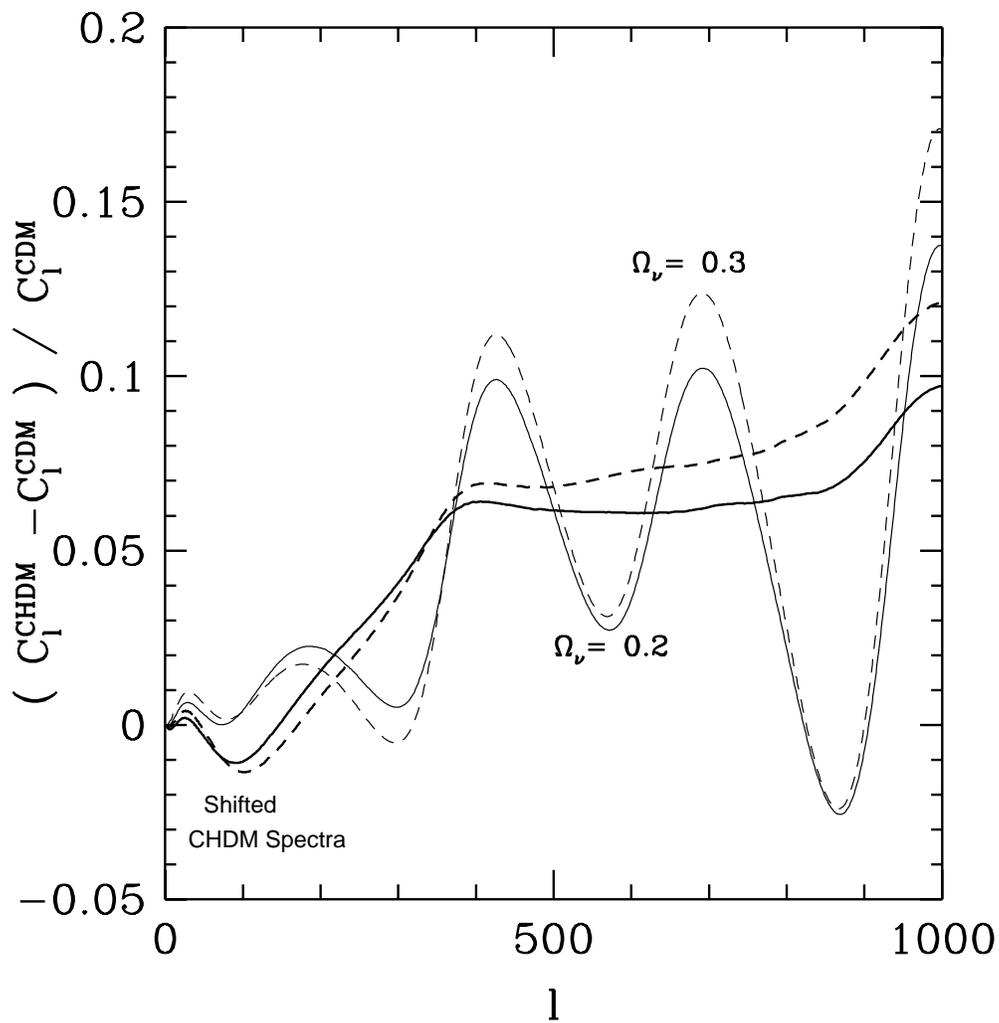

Fig. 5.— The fractional difference between the $C_l$'s in CHDM and CDM. The difference oscillates due to the sound horizon shift. If this is removed by shifting the CHDM spectra in $l$–space, the difference remains constant in the range $400 \lesssim l \lesssim 900$.

Newtonian potential (e.g. see Kodama and Sasaki (1984)). Combining equation 13 with the photon continuity equation we can obtain the solutions $\Theta_0$, $\Theta_1$ at recombination. These solutions are then evolved to the present with simple freestreaming formulae. In particular, there are three contributions to $\Theta_l$ today:

$$\Theta_l^{\text{Monopole}}(\eta_0) = [\Theta_0 + \Psi](\eta_*)j_l\left(k(\eta_0 - \eta_*)\right) \tag{14}$$

$$\Theta_l^{\text{Dipole}}(\eta_0) = 3\Theta_1(\eta_*)\left[\frac{l j_{l-1}\left(k(\eta_0 - \eta_*)\right) - (l+1)j_{l+1}\left(k(\eta_0 - \eta_*)\right)}{2l+1}\right] \tag{15}$$

$$\Theta_l^{\text{Integrated Sachs-Wolfe}}(\eta_0) = \int_{\eta_*}^{\eta_0} d\eta \left[\dot{\Psi} - \dot{\Phi}\right] j_l\left(k(\eta_0 - \eta)\right) \tag{16}$$

where the $j_l$'s are spherical Bessel functions; the $k$ dependence everywhere has been suppressed; and $\eta_*$ is the conformal time at decoupling. To get the $C_l$'s one simply adds these three contributions and then integrates over $k$:

$$C_l = \frac{2}{\pi} \int_0^\infty \frac{dk}{k} \, k^3 |\Theta_l(\eta_0)|^2. \tag{17}$$

To get a more physical understanding of the difference between the CHDM and CDM spectra, we examine the separate contributions of each of the three terms in equations 14-16 below.

### 3.4. Monopole and Dipole

We have extracted from our Boltzmann code the various terms in equations 14-16. The top panel of figure 6 shows the sum $\Theta_0(\eta_*) + \Psi(\eta_*)$ (the monopole term) as a function of $k$ in the two models. The striking feature of this graph is the point made by HS: the peaks in the monopole here line up exactly with the peaks in the $C_l$'s shown in figure 2. Thus the monopole is the dominant contribution at small scales; it is the first place to look in our search for the difference between the spectra of CDM and CHDM. And, in fact, the CHDM monopole is a few percent higher at the first peak and $\sim 10\%$ higher at the next two peaks. As can be seen from the bottom panel of figure 6, this difference in the monopole amplitudes translates directly into a several percent difference in $C_l$ for $l \gtrsim 200$. Figure 6 also shows the dipoles in the two models. We see that the dipoles, while smaller, also exhibit the same tendency towards higher CHDM amplitudes at the peaks. The bottom panel illustrates how this shows up in the $C_l$'s. Adding these two incoherently leads to our semi-analytic estimate for the difference: for $l \gtrsim 200$, this estimate gives excellent agreement with the numerical work.

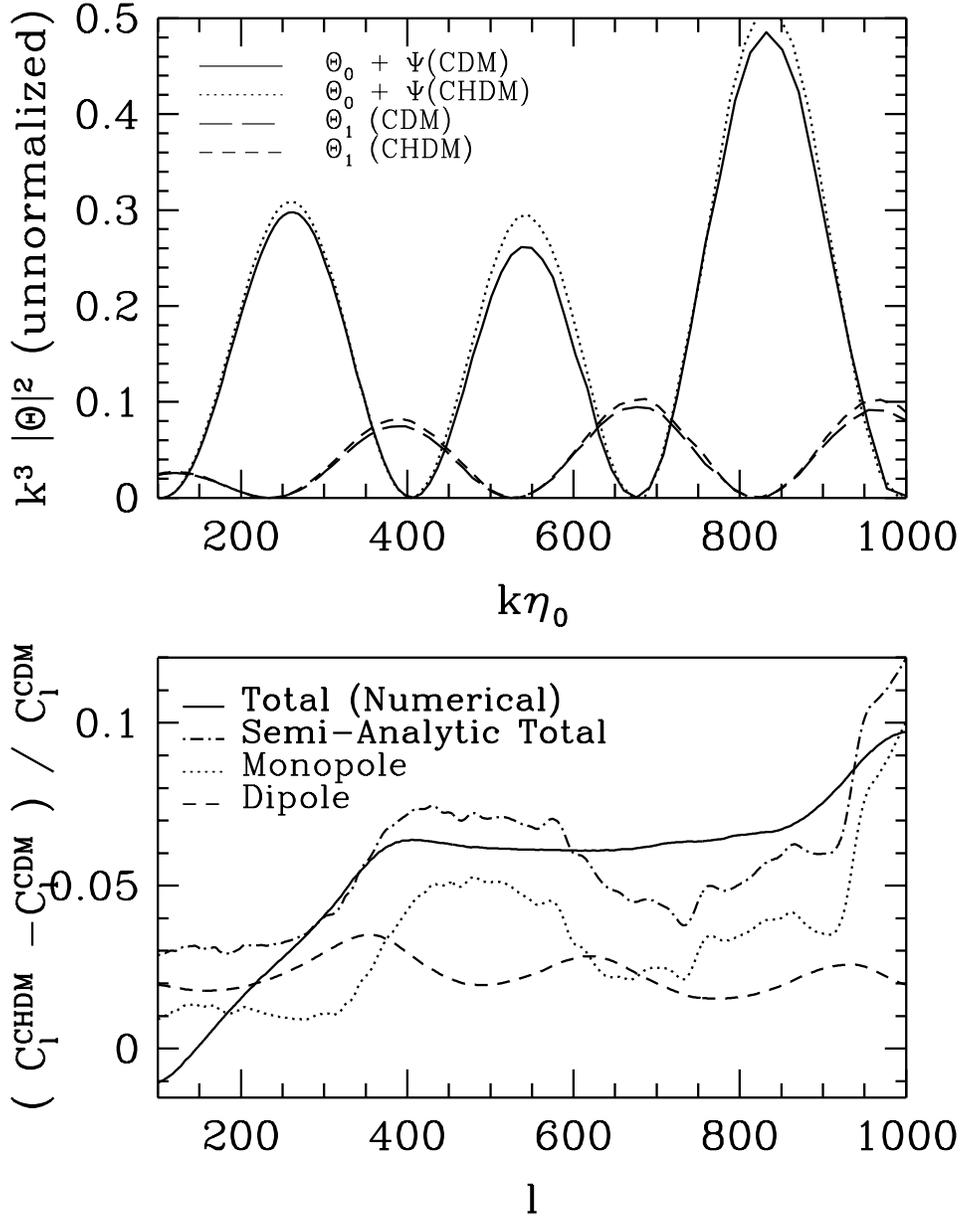

Fig. 6.— *Top Panel.* The monopole and dipole at decoupling. As shown by HS the peaks in the monopole (at $k\eta_0 \sim 200, 500, 800$) correspond to the peaks in the $C_l$'s (at comparable values of $L$). The amplitudes of the peaks are higher in CHDM than in CHDM (here $\Omega_\nu = 0.2$). The second peak (at $k\eta_0 \sim 500$) is $\sim 10\%$ higher in CHDM. The peaks of the dipole exhibit the same tendency. *Bottom Panel.* The resulting difference in the $C_l$'s due to the monopole and dipole. Also shown is the exact numerical result. Note the excellent agreement for $l \gtrsim 200$. All CHDM spectra in both panels have been shifted to account for the difference in sound horizons.

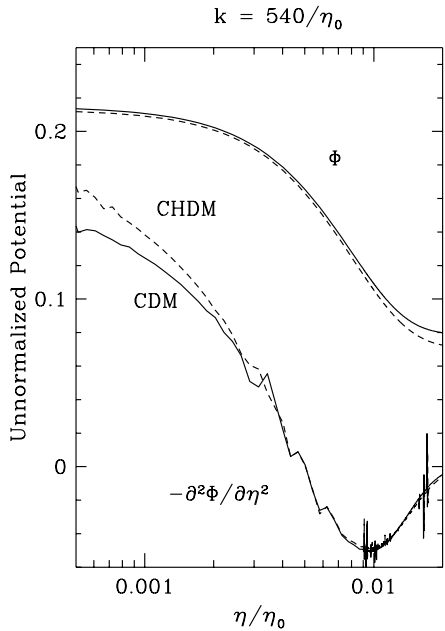

Fig. 7.— The potential $\Phi$ and its second derivative in CDM and CHDM ($\Omega_\nu = 0.2$). $\Phi$ decays in CHDM due to neutrino freestreaming, so the second derivative has a larger amplitude. The larger $-\ddot{\Phi}$ means the forcing function for the monopole is larger in CHDM than in CDM.

The major contribution to the increase in the monopole term at last scattering can be traced to an increase in the forcing function (equation 13), and in particular to an increase in the $\ddot{\Phi}$ component – neutrino freestreaming on these scales causes a larger decay in the potential. In CHDM models, anisotropies on scales smaller than the horizon size when the (massive) neutrinos become non-relativistic will receive an additional push from the increase in $\ddot{\Phi}$. In Figure 7 we show $\ddot{\Phi}$ for $k\eta_0 = 540$, which corresponds approximately to the location of the second Doppler peak.

The only place where the monopole + dipole do not fully explain the difference is for $l \lesssim 300$. Again, the difference in the expansion law and the corresponding change in the gravitational potentials are responsible for most of the difference, as we discuss below.

### 3.5. Integrated Sachs-Wolfe Effect

If the gravitational potentials $(\Phi, \Psi)$ do not remain constant after the time of last scattering of the photons, then there is an additional contribution to the anisotropy spectrum, the integrated Sachs-Wolfe effect (ISW). Although the potentials *do* remain relatively constant in many cosmological models, including the ones we are considering here, HS have shown that even small changes in the potential lead to large effects in the $C_l$'s. To understand this, HS suggested a simple approximation to the ISW integral of equation 16. They point out that $\dot{\Phi}$ and $\dot{\Psi}$ are typically non-zero only at early times. Therefore the major contribution to the ISW integral comes from $\eta \simeq \eta_\star$. A reasonable approximation then is to set the argument of the Bessel function to $k(\eta_0 - \eta_\star)$, thus reducing the ISW term to

$$\Theta_l^{\text{Integrated Sachs-Wolfe}}(\eta_0) \simeq \left[\Psi(\eta_0) - \Psi(\eta_\star) - \Phi(\eta_0) + \Phi(\eta_\star)\right] \, j_l\left(k(\eta_0 - \eta_\star)\right). \quad (18)$$

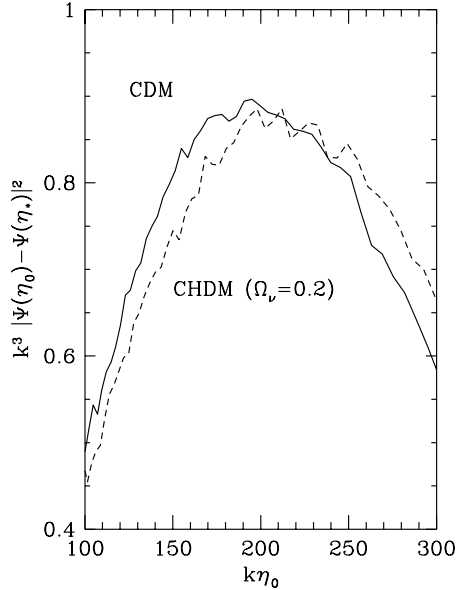

Fig. 8.— The change in the gravitational potential (unnormalized) from recombination to today as a function of scale in CDM and CHDM ($\Omega_\nu = 0.2$). At low $k$ ($\lesssim 200/\eta_0$), the CDM potential changes slightly more; this leads to a larger ISW effect at these relevant scales. At smaller scales, the ISW effect ceases to be important since it no longer contributes in phase with the monopole.

This approximation breaks down on small scales, but it does have one virtue: it shows clearly why the ISW effect plays such a crucial role in understanding the spectrum of anisotropies. The key point is that, with this approximation, the ISW effect is now seen to add coherently to the monopole: they are both proportional to the spherical Bessel function, $j_l\left(k(\eta_0 - \eta_\star)\right)$. The magnitude of the effect in this approximation is proportional to the difference between the values of the potential today and at last scattering. Figure 8 plots this difference in both CDM and CHDM. At low $k$, where the approximation is valid, the potential changes slightly more in CDM. The reason for this can be seen in Fig. 9, which shows the potential as a function of $\eta$ for the two models for $k\eta_0 = 150$. Taking into account the fact that $\eta_*^{\mathrm{CHDM}} > \eta_*^{\mathrm{CDM}}$, it is clear that the change in $\Psi$ between $\eta_*$ and $\eta_0$ is larger in CDM on this scale and of the right magnitude to be consistent with Figure 8. Thus we expect a larger ISW effect in CDM than in CHDM.

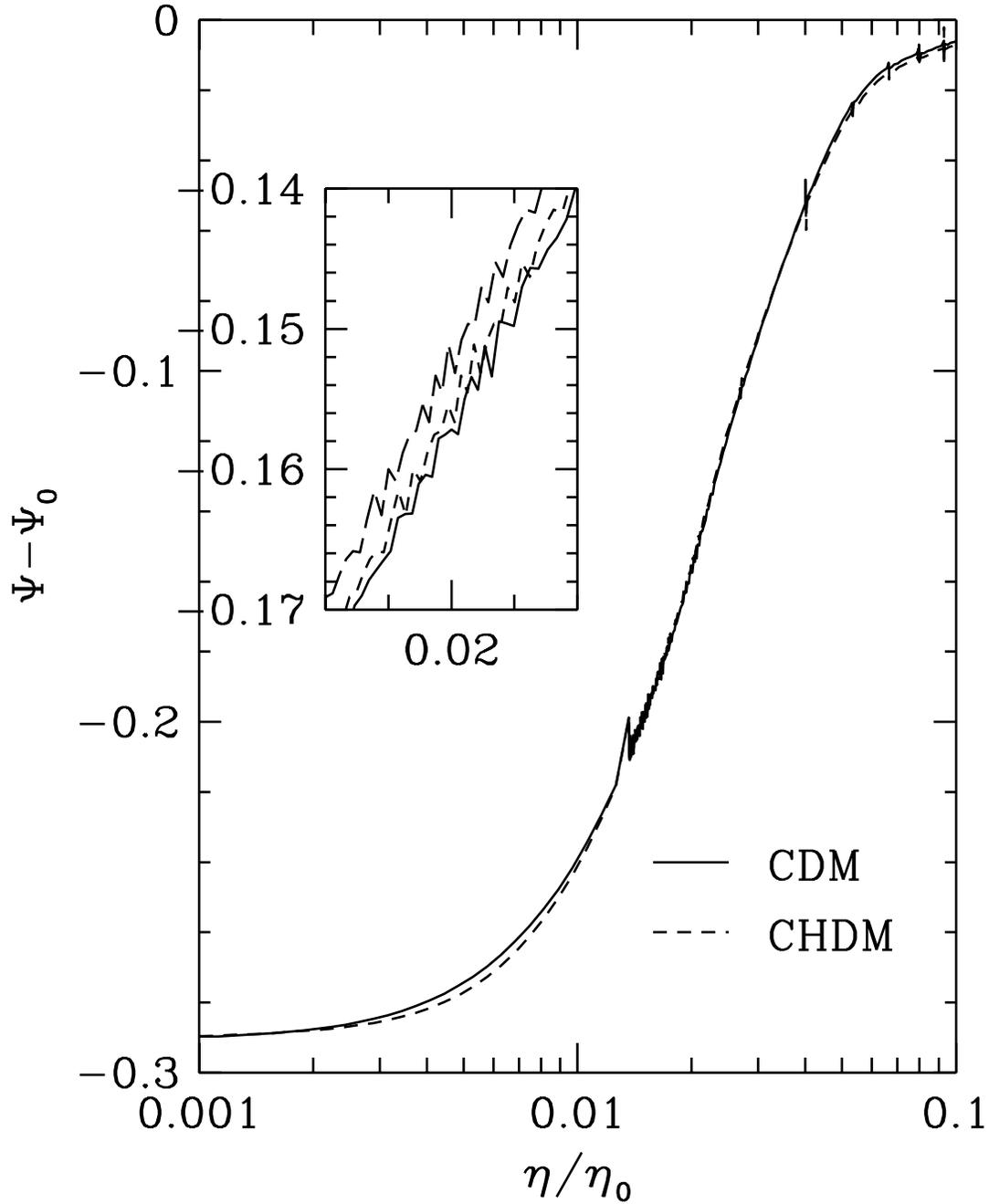

Fig. 9.— The potential for $k\eta_0 = 150$ in both CDM and CHDM as a function of conformal time. The region around recombination is blown up. The upper line in the blown up region is CHDM but with $\eta$ shifted to account for the later recombination $\eta$ in CHDM. Since the potential is changing very rapidly in this region, even a small change in the recombination time causes a large change in the ISW effect.

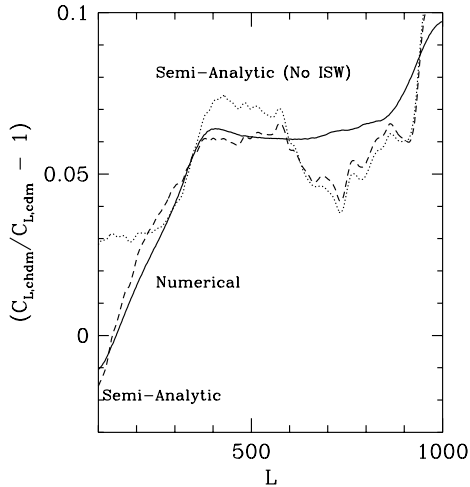

Fig. 10.— The sources of the difference between CDM and CHDM ($\Omega_\nu = 0.2$) (Again CHDM has been shifted in $l$–space to account for the different sound speeds. The monopole and the dipole (dotted line labeled "Semi-Analytic (No ISW)") account for the bulk of the difference for $l \gtrsim 300$. At smaller $l$, the ISW effect needs to be included as well since it adds coherently to the monopole.

We can summarize the effects discussed in this section and the preceding section as follows. The largest contribution to the difference in amplitude of the CDM and CHDM spectra is due to an increase in the monopole term at last-scattering in CHDM models, and is important on small scales ($l \gtrsim 400$). This can be attributed to neutrino freestreaming, which increases the decay of the potential on these scales and thus gives an additional boost to the amplitude of the acoustic oscillations via the forcing function. On larger scales, there is a smaller effect due to the change in the expansion law. Neutrino freestreaming is no longer important, but because the epoch of matter-radiation equality has been delayed in the case of CHDM, the change in the gravitational potential on these scales is shifted slightly to later times in CHDM models and last-scattering occurs at a later conformal time (see for example fig. 9). This results in a larger ISW effect after recombination for CDM models.

We can now combine all the results of this section into figure 10. Shown is the difference between CDM and (shifted) CHDM. Without the ISW effect, the semi-analytic approach does a good job of accounting for the differences for $l \gtrsim 300$. Including the ISW effect leads to excellent agreement for the full range of $l$.

## 4.  Can Experiments Distinguish CDM from CHDM?

The goal of the next generation of satellite experiments is to produce an all-sky map with angular resolution of order half a degree or better. To determine whether this type of experiment will be able to distinguish CDM from CHDM, we must first calculate how accurately the $C_l$'s will be determined by such an experiment. A very useful formula has been derived by Knox (1995) for the experimental uncertainty $\Delta C_l$:

$$\frac{\Delta C_l}{C_l} = \sqrt{\frac{2}{2l+1}} \left( 1 + \frac{\sigma_{\mathrm{pixel}}^2 \Omega_{\mathrm{pixel}}}{C_l} \exp\{l^2 \sigma_{\mathrm{beam}}^2\} \right) \qquad (19)$$

where $\sigma_{\mathrm{beam}} = .425\theta_{\mathrm{fwhm}}$ is the beam size; $\sigma_{\mathrm{pixel}}$ is the noise per pixel; and $\Omega_{\mathrm{pixel}}$ is the area per pixel. Knox noted that, for fixed observing time, the product $\sigma_{\mathrm{pixel}}^2 \Omega_{\mathrm{pixel}}$ remains constant as the beam size changes. In what follows, we will be varying the beam size to see how this affects our ability to distinguish CDM from CHDM. As we vary the beam size, though, we will keep $\sigma_{\mathrm{pixel}}^2 \Omega_{\mathrm{pixel}}$ fixed at $(44\mu K)^2(20' \times 20')$. This noise level is certainly attainable by the next generation of experiments, even accounting for the noise due to foregrounds.

Equation 19 shows that even if the noise is very low, there is still an unavoidable uncertainty $\Delta C_l/C_l = \sqrt{2/(2l+1)}$. This minimum uncertainty, dubbed *cosmic variance*, results from the fact that in most theories, the observed $a_{lm}$'s are drawn from a distribution with variance $C_l$. To know the true variance exactly, one would have to sample the distribution an infinite number of times. In the real world, this is impossible, as we only get $2l + 1$ chances for each $l$. Hence the unavoidable $\sqrt{2/(2l+1)}$ uncertainty.

Figure 11 shows $\Delta C_l/C_l$ for a variety of beam widths. Experiments cannot resolve features smaller than the beam size, so the uncertainty in $C_l$ becomes large at large $l$ (note the exponential factor in equation 19). Nonetheless, we can expect to obtain information about the $C_l$'s out to $l \sim 500$ and perhaps even further with smaller beam sizes.

The information about $C_l$'s out to $l \gtrsim 500$ can be used to discriminate CDM from CHDM. To see this, let us suppose that the only difference between CDM and CHDM was a 10% shift in the $C_l$'s between $l = 400$ and $l = 500$. From Figure 11 we see that for $\theta_{\mathrm{fwhm}} = 30'$, $\Delta C_l/C_l \sim 0.5$ over this range. Thus, using any one value of $l$, one would not be able to distinguish CDM from CHDM. However, using all hundred values of $l$ in this range one beats down the uncertainty by a factor of ten ($\sqrt{100}$). Thus a 10% shift would be detectable. This crude argument, together with figure 11, suggests that beam widths larger than $\theta_{\mathrm{fwhm}} = 30'$ would be unable to distinguish CDM from CHDM.

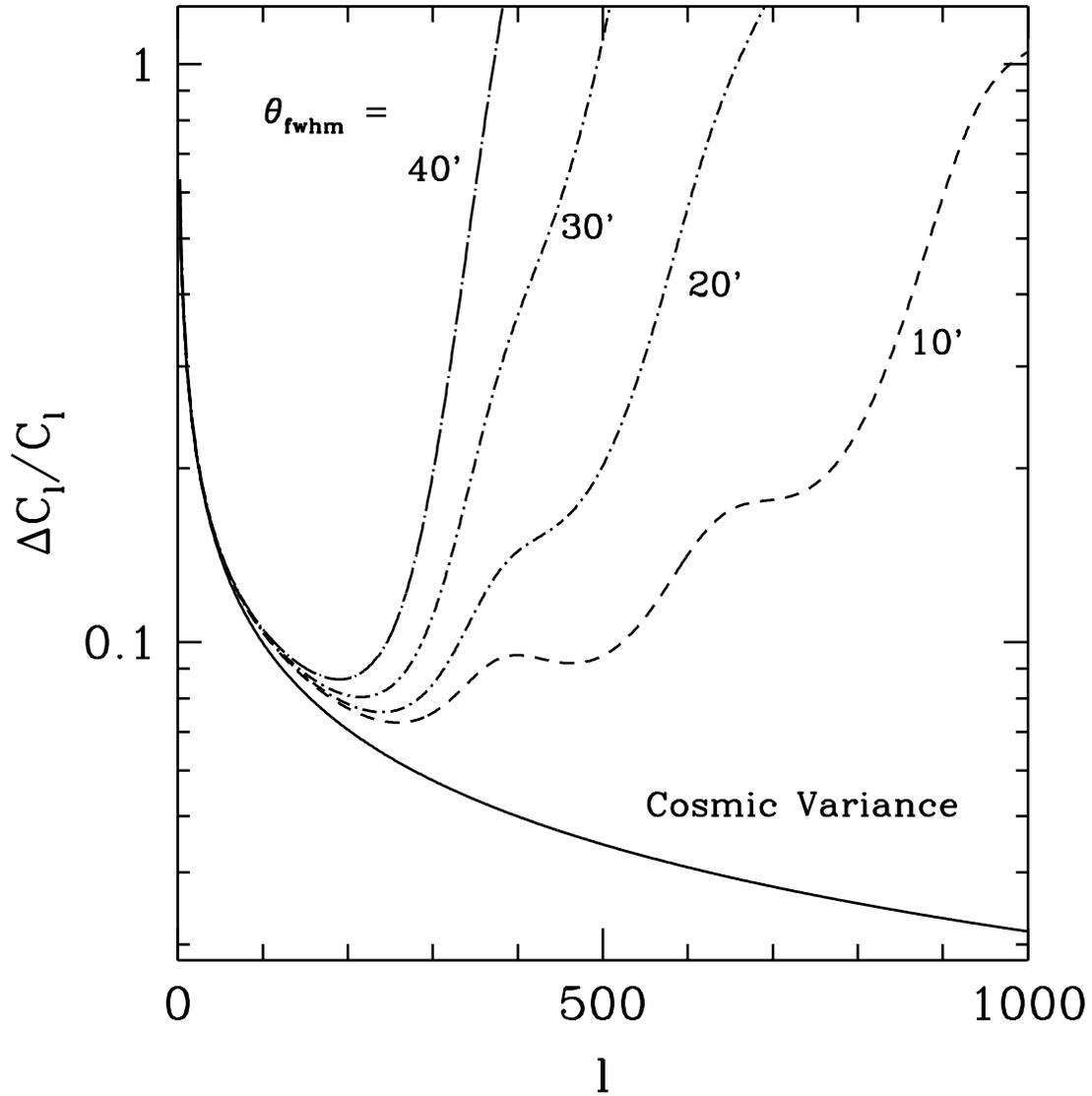

Fig. 11.— The uncertainty in the measured $C_l$'s from an all-sky experiment with $\sigma^2_{\rm pixel}\Omega_{\rm pixel}$ fixed at $(44\mu K)^2 (20' \times 20')$. The cosmic variance line represents the minimum possible uncertainty. Here the "true" $C_l$'s were taken to be the standard CDM spectrum.

Let us make the argument slightly more quantitative, following Jungman et al. Given experimentally observed $C_l^{exp}$ with $\Delta C_l$'s, we can write down a goodness-of-fit function

$$\chi^2(\Omega_\nu) = \sum_l \frac{(C_l^{exp} - C_l^{th}(\Omega_\nu))^2}{(\Delta C_l)^2}. \qquad (20)$$

Here, we are supposing that we know all other cosmological parameters (baryon density, Hubble parameter, cosmological constant, initial perturbation spectrum, ...) so that the theoretical prediction $C_l^{th}$ depends only on the energy density of massive neutrinos, $\Omega_\nu$. Presumably the value of $\chi^2$ from a given experiment will have its minimum at a value of $\Omega_\nu$ pretty close to the true value. So the question of how well we can determine $\Omega_\nu$ boils down to the question of how fast does $\chi^2$ change as we change $\Omega_\nu$ away from the true value. That is, we are interested in the behavior of the $\chi^2$ function near its minimum. Thus it makes sense to expand:

$$\chi^2(\Omega_\nu) = \chi^2(\overline{\Omega}_\nu) + \frac{1}{2}\frac{\partial^2\chi^2}{d\Omega_\nu^2}|_{\Omega_\nu=\overline{\Omega}_\nu} \ (\Omega_\nu - \overline{\Omega}_\nu)^2 + \dots \qquad (21)$$

Here $\overline{\Omega}_\nu$ is the value of $\Omega_\nu$ which minimizes the $\chi^2$ (thus there is no first derivative term in equation 21). With some mild assumptions detailed in Press et al. (1986), the one-$\sigma$ error on the parameter $\Omega_\nu$ is then determined by the coefficient of the quadratic term in equation 21:

$$\langle(\Omega_\nu - \overline{\Omega}_\nu)^2\rangle = \left[\frac{1}{2}\frac{\partial^2\chi^2}{d\Omega_\nu^2}|_{\Omega_\nu=\overline{\Omega}_\nu}\right]^{-1} \qquad (22)$$

To convincingly discriminate CDM from CHDM, this one-$\sigma$ error should be much smaller than 0.2 (the preferred value of $\Omega_\nu$ in CHDM). Figure 12 shows this error as a function of beam width, again under the unrealistic assumption that all other parameters are known. (The assumed true spectrum here is CDM – $\Omega_\nu = 0$.) As expected, beam widths greater than $\sim 30'$ cannot discriminate CDM from CHDM.

In the real world, of course, there are bound to be other unknowns besides $\Omega_\nu$. equation 22 is easily extended to account for this. Both sides become matrices, with each element in a row/column corresponding to a different variable. The one-$\sigma$ error on a given variable is given by the analogue of equation 22 with the appropriate indices attached. This is equivalent to integrating over all other variables [i.e. allowing them to take any value] and finding the uncertainty in the one remaining parameter of interest.

As an example, we first integrate over the normalization, $Q$. Allowing for this unknown leads to the dot-dashed curve in figure 12. Note that this completely removes any discriminatory power from experiments with $\theta_{fwhm} \gtrsim 30'$. (This ability in the previous curve stemmed from the very small differences between CDM and CHDM at the first Doppler peak.) Another parameter which at present is not known very well is $n$, the spectral index of the primordial perturbations. Allowing $n$ to vary leads to the long dashed curve in figure 12. It is hard to confuse a change in $n$ with a change in $\Omega_\nu$, so allowing $n$ to vary doesn't change things much. Allowing the baryon density $\Omega_b h^2$ and the Hubble constant $h$ to vary, though, does make things significantly worse.

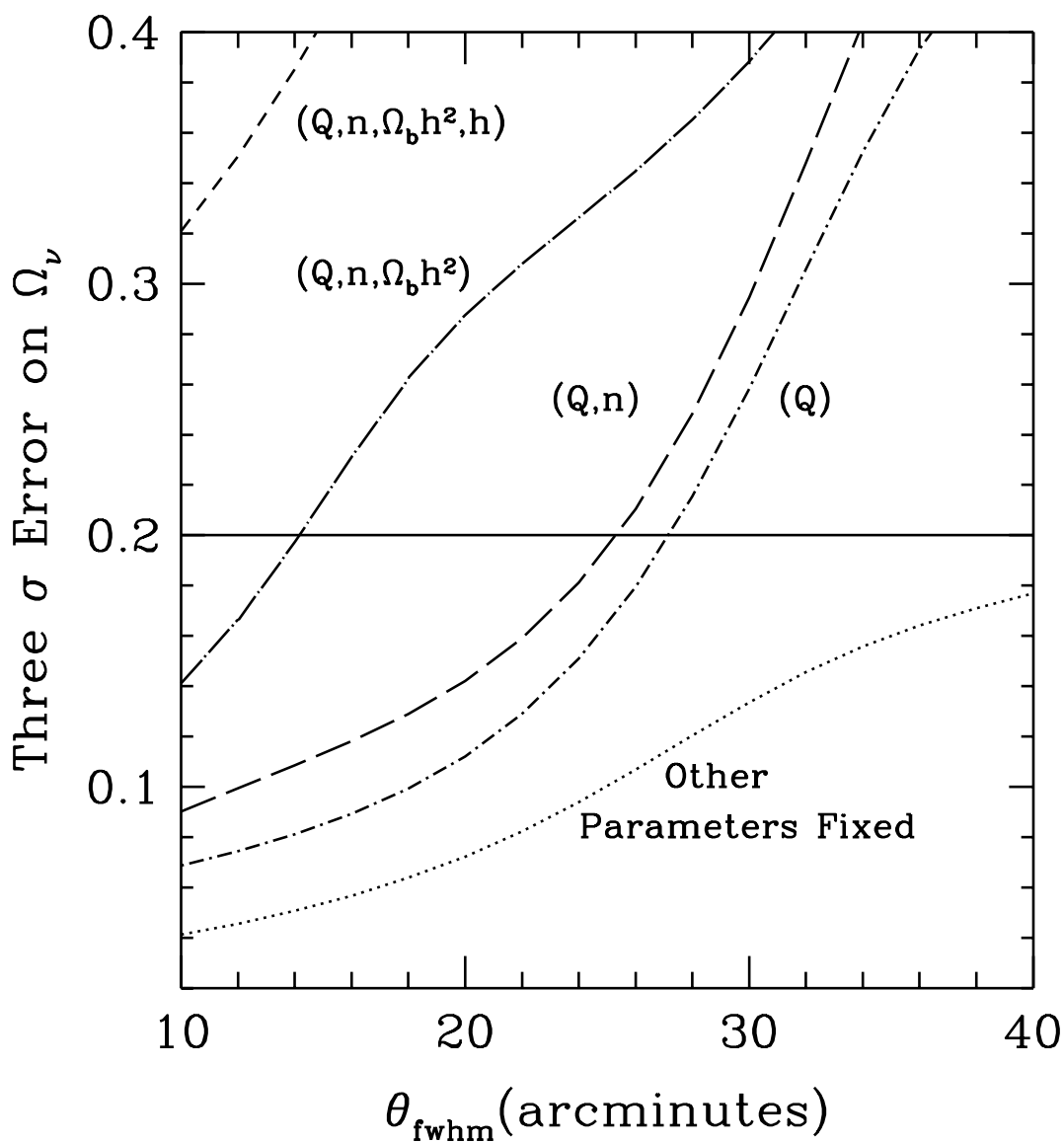

Fig. 12.— The 3-$\sigma$ error on massive neutrino energy density. Lowest curve assumes all other parameters are known. Each successive curve adds an additional unknown parameter. The horizontal solid line represents the preferred value of $\Omega_\nu$ in CHDM. Curves below this succeed in discriminating CDM from CHDM.

This preliminary investigation suggests that there will be a possibility of determining $\Omega_\nu$ from CMB measurements. Accounting for other free parameters worsens the situation but not beyond reason.

We thank Wayne Hu and Marc Kamionkowski for helpful conversations. This work was supported in part by the DOE (at Chicago and Fermilab) and the NASA (at Fermilab through grant NAG 5-2788).